\documentclass[a4paper,11pt]{article}
\usepackage{lmodern}

\usepackage[T1]{fontenc}
\usepackage[latin9]{inputenc}
\usepackage{mathrsfs}
\usepackage{amsmath}
\usepackage{graphicx}

\makeatletter
\newcommand{\lyxaddress}[1]{
	\par {\raggedright #1
		\vspace{1.4em}
		\noindent\par}
}

\usepackage{mathrsfs}   %\mathscr{L}
\usepackage{slashed}     %\slashed{p}
\usepackage{bbold}  % \mathbb{1} for the identity matrix
\usepackage{url}
\usepackage{graphicx}
\usepackage[colorlinks=true,linkcolor=redLinks,citecolor=greenLinks,urlcolor=redLinks, pdfborder={0 0 1}]{hyperref}
\usepackage{xcolor}
\usepackage{framed}
\usepackage[numbers,sort&compress]{natbib}
\usepackage{amsmath}

\allowdisplaybreaks

\colorlet{shadecolor}{gray!15}

\definecolor{greenLinks}{rgb}{0, 0.6, 0} 
\definecolor{blueLinks}{rgb}{0, 0, 0.6}
\definecolor{redLinks}{rgb}{0.6, 0, 0}
\definecolor{tempText}{rgb}{0.55, 0.10,0.67}
\definecolor{eprintLinks}{rgb}{0.4, 0.4, 0.4}
\definecolor{journalLinks}{rgb}{0.6, 0, 0}

\newcommand{\MYhref}[3][redLinks]{\href{#2}{\color{#1}{#3}}}%

\usepackage{multirow}
\let\orig@Hy@EveryPageAnchor\Hy@EveryPageAnchor
\def\Hy@EveryPageAnchor{%
	\begingroup
	\hypersetup{pdfview=Fit}%
	\orig@Hy@EveryPageAnchor
	\endgroup
}

\let\oldFootnote\footnote
\newcommand\nextToken\relax

\renewcommand\footnote[1]{%
	\oldFootnote{#1}\futurelet\nextToken\isFootnote}

\newcommand\isFootnote{%
	\ifx\footnote\nextToken\textsuperscript{,}\fi}

\definecolor{myPurple}{RGB}{128,0,182}

\makeatother

\begin{document}
	\title{Explaining the SM flavor structure with grand unified theories}
	\author{Renato M. Fonseca\date{}\footnote{Contribution to the proceedings of the 40th International Conference on High Energy physics (ICHEP2020), 28 July -- 6 August 2020, Prague, Czech Republic (virtual meeting).}}
	\maketitle
	
	\lyxaddress{\begin{center}
			{\Large{}\vspace{-0.5cm}}Institute of Particle and Nuclear Physics\\
			Faculty of Mathematics and Physics, Charles University,\\
			V Holešovi\v{c}kách 2, 18000 Prague 8, Czech Republic\\
			~\\
			Email: fonseca@ipnp.mff.cuni.cz
			\par\end{center}}
	\begin{abstract}
		We do not know why there are three fermion families in the Standard
		Model (SM), nor can we explain the observed pattern of fermion masses
		and mixing angles. Standard grand unified theories based on the $SU(5)$
		and $SO(10)$ groups fail to shed light on this issue, since they
		also contain three copies of fermion representations of an enlarged
		gauge group. However, it does not need to be so: the Standard Model
		families might be distributed over distinct representations of a grand
		unified model, in which case the gauge symmetry itself might discriminate
		the various families and explain (at least partially) the flavor puzzle.
		The most ambitious version of this idea consists on embedding all
		SM fermions in a single irreducible representation of the gauge group.
	\end{abstract}

\section{Flavor and grand unification}

The Standard Model of particle physics is a gauge theory built around
the group $SU(3)\times SU(2)\times U(1)\equiv G_{SM}$, with the fermions
distributed over 15 irreducible representations: $3(Q+u^{c}+d^{c}+L+e^{c})$.
The quantum numbers $Q=\left(\boldsymbol{3},\boldsymbol{2},1/6\right)$,
$u^{c}=\left(\overline{\boldsymbol{3}},\boldsymbol{1},-2/3\right)$,
$d^{c}=\left(\overline{\boldsymbol{3}},\boldsymbol{1},1/3\right)$,
$L=\left(\boldsymbol{1},\boldsymbol{2},-1/2\right)$ and $e^{c}=\left(\boldsymbol{1},\boldsymbol{1},1\right)$
are somewhat curious; for example, only the fundamental and anti-fundamental
representations of the special unitary groups are used, plus it is
possible to write all hypercharges as rational numbers. Grand unified
theories (GUTs) provide an elegant explanation for these quantum numbers.
The idea is that the true gauge symmetry of Nature is given by a group
$G$ larger than $G_{SM}$, which is spontaneously broken at very
high energies. If $G$ is a simple group --- such as $SU(5)$, $SO(10)$
or $E_{6}$ \cite{Georgi:1974sy,Georgi:1974my,Fritzsch:1974nn,Gursey:1975ki} ---
it then becomes possible to relate the three Standard Model gauge
coupling constants. Furthermore, the quantum numbers of fermions under
$G_{SM}$ follow directly from the transformation properties of these
fields under the enlarged group $G$. %For example, the representation
%$\overline{\boldsymbol{5}}$ of $SU(5)$ decomposes under the $G_{SM}$
%subgroup as $d^{c}+L$ , and the $\boldsymbol{10}$ contains $Q+u^{c}+e^{c}$.
%On the other hand, the spinor representation of $SO(10)$ decomposes
%as $\boldsymbol{16}\rightarrow Q+u^{c}+d^{c}+L+e^{c}+N^{c}$, with
%$N^{c}=\left(\boldsymbol{1},\boldsymbol{1},0\right)$ having the properties
%of a sterile neutrino.

Given that grand unified theories can account for the quantum numbers
of the Standard Model fermions in an appealing way, we are left with
what seems to be a deeper mystery of the Standard Model, namely the
existence of three copies of every irreducible fermion representation
$X$ of $G_{SM}$ ($X=Q$, $u^{c}$, $d^{c}$, $L$ and $e^{c}$).
Each copy is often called a \textit{family}, a \textit{generation}
or a \textit{flavor}. Assigning a flavor index to each fermion, Yukawa
interactions are controlled by $3\times3$ matrices
which account for the measured fermion masses and mixing parameters:
\begin{equation}
\left(Y_{U}\right)_{ij}Q_{i}u_{j}^{c}H+\left(Y_{D}\right)_{ij}Q_{i}d_{j}^{c}H^{*}+\left(Y_{E}\right)_{ij}L_{i}e_{j}^{c}H^{*}
\end{equation}
It is worth noting that in standard GUTs --- based on the groups
mentioned earlier --- the fermion representations
also have flavor indices and therefore at high energies the Yukawa
couplings can still be seen as matrices in flavor space.

We also do not know why these matrices have the values that they
do, but perhaps once we have an explanation for the existence of three
families that might become clear. One possibility is that the Standard
Model family replication is an accident in the following sense: under
a more fundamental gauge group, fermions might be assigned to a combination
of representations $R+R^{\prime}+\cdots$ which do not have a trifold
repetition. This is in fact what happens with several (but not all)
models based on the semi-simple group $SU(3)\times SU(3)\times U(1)$
\cite{Singer:1980sw,Pisano:1991ee,Frampton:1992wt,Fonseca:2016tbn}.
However, it can also happen with simple groups. A particularly interesting
example is the $SU(11)$ model \cite{Georgi:1979md,Fonseca:2015aoa}
where fermions are assigned to the representations $\overline{\boldsymbol{11}}+\overline{\boldsymbol{55}}+\overline{\boldsymbol{165}}+\boldsymbol{330}$:
family replication is nowhere to be seen at a fundamental level, only
emerging at low energies due to spontaneous symmetry breaking. 

\begin{figure}[tbph]
\begin{centering}
\includegraphics[scale=0.32]{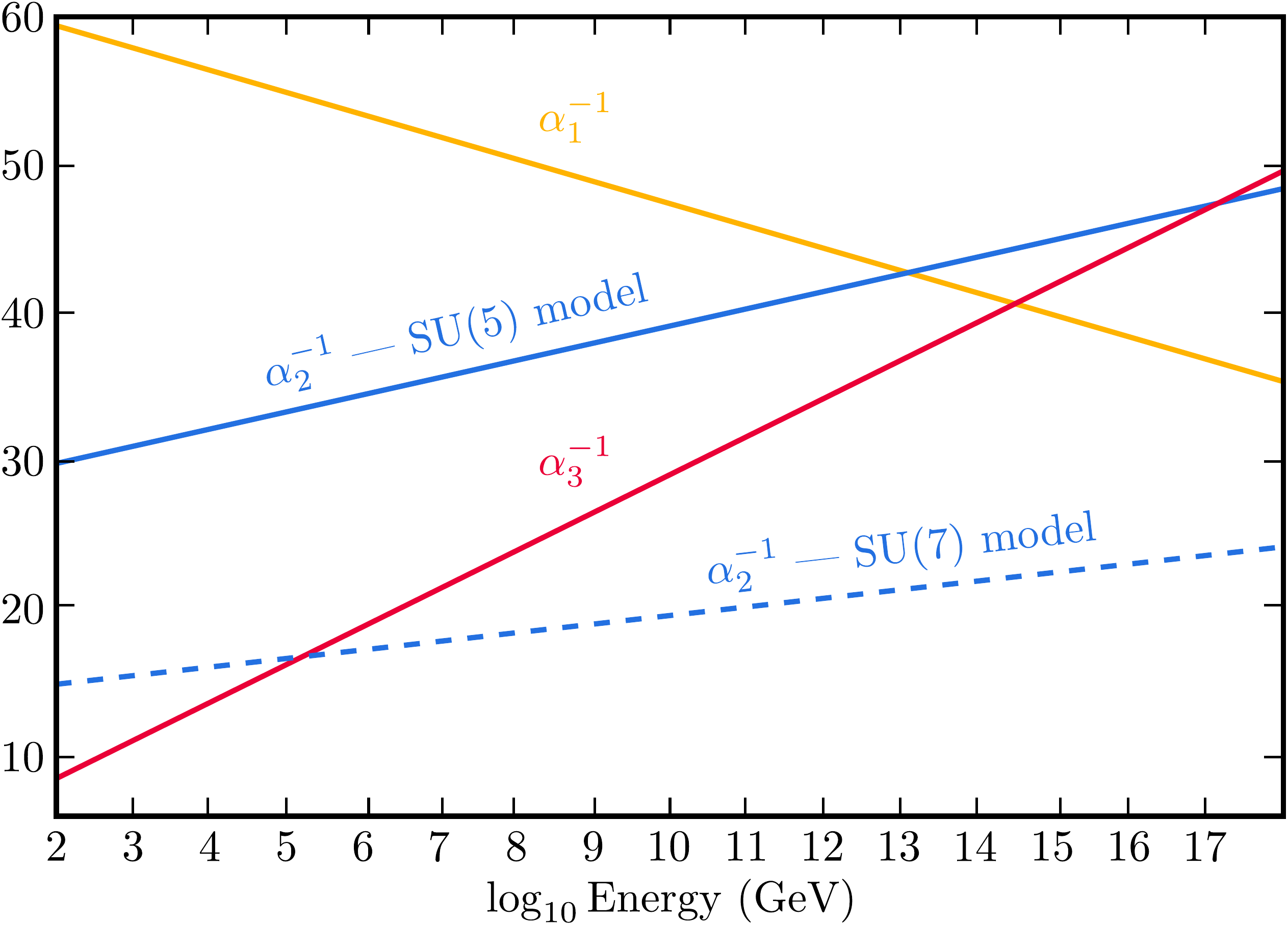}
\par\end{centering}
\caption{\label{fig:1}Bottom-up evolution of the Standard Model gauge couplings
at one loop, with $\alpha_{i}^{-1}\equiv4\pi/g_{i}^{2}$.
The usual picture (full lines) makes the assumption that fermions
are embedded as in $SU(5)$. But other scenarios might spoil the relation
$g_{1}=\sqrt{5/3}g^{\prime}$ and it is even possible that
$\left(g_{2},g_{3}\right)\protect\neq\left(g,g_{s}\right)$: for example
an $SU(7)$ embedding might yield the dashed line.}
\end{figure}

This constitutes a strong motivation for studying viable ways of embedding
the Standard Model fermions in GUT representations beyond the usual
scenarios. There are other reasons: for example, the way SM fermions
are embedded can have a rather curiously and dramatic effect on the
unification of the three gauge couplings. To appreciate it, let us
consider first the standard extrapolation of the value of these couplings
to higher energies, assuming only the Standard Model fields (full
lines in figure \ref{fig:1}), which are known not to unify. Importantly,
one usually compares the values of $g_{1}=\sqrt{5/3}g^{\prime}$,
$g_{2}=g$ and $g_{3}=g_{s}$, with the factor $\sqrt{5/3}$ being directly
related to the way in which fermions are presumed to be embedded.
In other words, figure \ref{fig:1} is not just a product of what
is known at low energies; it also incorporates an assumption. The
assumption is that the $d^{c}$ and the $L$ fermions are contained
in the representation $\overline{\boldsymbol{5}}$ of $SU(5)$. In
particular, the hypercharge matrix $Y=n\,\textrm{diag\ensuremath{\left(1/3,1/3,1/3,-1/2,-1/2\right)}}$
must have the same norm as all other $SU(5)$ generators, so equating
$\textrm{Tr}\left(YY\right)$ with $\textrm{Tr}\left(T_{SU(2)}^{3}T_{SU(2)}^{3}\right)$
--- where $T_{SU(2)}^{3}=\textrm{diag}\left(0,0,0,1/2,-1/2\right)$ is the diagonal
SM $SU(2)$ generator --- yields $n=\sqrt{3/5}$,
therefore $y_{\textrm{norm.}}=\sqrt{3/5}y$ and $g_{1}=\sqrt{5/3}g^{\prime}$.

This factor is the same for $SO(10)$ and $E_{6}$ GUTs, but different
arrangements could conceivably yield $n\neq\sqrt{3/5}$, potentially
leading to a situation where the three gauge couplings unify with
no extra fields lighter than the unification scale. This is certainly
possible mathematically; the question is whether or not those scenarios
are associated with viable models. In fact, mathematically one can
conceivably even spoil the relations $g_{2}=g$ and $g_{3}=g_{s}$.
For example, an $SU(7)$ model utilizing the branching rule $\boldsymbol{7}\rightarrow d^{c}+L+\left(\boldsymbol{1},\boldsymbol{2},0\right)$
would imply that $g_{2}=\sqrt{2}g$ (dashed line in figure \ref{fig:1}).\footnote{One might ask why is it so important to normalize all generators of
the GUT group such that $\textrm{Tr}\left(T^{a}T^{b}\right)\propto\delta^{ab}$.
The reason is this: the structure constants $c^{abc}$ appearing in
the commutator relation $\left[T^{a},T^{b}\right]=ic^{abc}T^{c}$ control
the transformation of gauge bosons under infinitesimal global transformations:
$A_{\mu}^{a}\rightarrow A_{\mu}^{a}-c^{bca}A_{\mu}^{b}\alpha^{c}$
(where $\alpha$ is the transformation parameter). And in order for
this to correspond to a unitary transformation, $c^{abc}$ must be
a completely antisymmetric tensor, which in turn requires that $\textrm{Tr}\left(T^{a}T^{b}\right)\propto\delta^{ab}$.}

\section{Viable fermion GUT representations}

The only fermions which can have a pure electroweak mass are most
likely the Standard Model ones, so if there are more fermions they
must have a mass even before electroweak symmetry is broken. These
extra fermions --- if they exist at all --- must be vector-like.
In order for this to be true, the fermion GUT representations $R+R^{\prime}+\cdots$
must decompose under $G_{SM}$ into the 3 chiral families plus a real
representation (of $G_{SM}$) which can be reducible:
\begin{equation}
R+R^{\prime}+\cdots\rightarrow3\left(Q+u^{c}+d^{c}+L+e^{c}\right)+\textrm{real representation}\,.
\end{equation}
%This real representation can be reducible: it could consist of one
%or more sterile neutrinos $N^{c}=\left(\boldsymbol{1},\boldsymbol{1},0\right)$
%and/or a fourth generation of vector-like charged leptons $E+E^{c}=\left(\boldsymbol{1},\boldsymbol{1},-1\right)+\left(\boldsymbol{1},\boldsymbol{1},1%\right)$,
%just to give some examples.

It might not sound like much, but this simple constraint on the fermion
field content of grand unified theories turns out to be quite stringent.
In reference \cite{Fonseca:2015aoa} a search was made over (a) different simple
groups $G$, (b) different combinations of representations $R$, $R^{\prime}$,
... of $G$ and (c) different embeddings of $G_{SM}$ in $G$ (there
might be more than one). Under some reasonable assumptions, such as
the nonexistence of confining interactions besides those of $SU(3)_{c}$
at very low energies, the scan showed that
\begin{itemize}
\item The viable simple groups $G$ are $SO(10)$, $E_{6}$ and the $SU(N)$'s
for $N\geq5$.
\item The Standard Model group can only be embedded in one way in all these
groups, except for $SU(N\geq15$). There is no viable alternative
to the relations $g_{1}=\sqrt{5/3}g^{\prime}$, $g_{2}=g$ and $g_{3}=g_{s}$.
\item Apart from trivial variations, the fermion content in $SU(5)$ ($3\times\overline{\boldsymbol{5}}+3\times\boldsymbol{10}$)
and $E_{6}$ ($3\times\boldsymbol{27}$) GUTs is unique. Furthermore,
for all practical purposes so is the one in an $SO(10)$ model ($3\times\boldsymbol{16}$).
On the other hand, for $SU(N\geq6)$ there are various ways of embeddings
the SM fermions, and crucially family replication is not a requirement.
\item The only non-trivial case where all fermions can be embedded in a
single representation\footnote{This can also trivially be achieved with the fundamental representation
of $SU(45)$, or even bigger special unitary groups if we were to consider
extra vector-like fermions.} is by using the $\boldsymbol{171}$ representation of $SU(19)$.
This possibility was also mentioned in \cite{Yamatsu:2018tnv}.
\end{itemize}
The main features of an $SU(19)$ model exploring this last idea were
considered in \cite{Ekstedt:2020gaj}, which provides a glimpse of how
flavor might arise from a fundamental theory which has no family
replication. I used the word \textit{glimpse} because, as we shall
see below, in order to compute the Standard Model Yukawa matrices
$Y_{U}$, $Y_{D}$ and $Y_{E}$ at the electroweak scale one needs
to calculate the ratio of several vacuum expectations values (VEVs),
which is a daunting task still to be addressed.

\section{A model for flavorgenesis}

With a single fermion representation $\boldsymbol{171}$ of the $SU(19)$
gauge group, the fermion-fermion-scalar interactions are controlled
by a singlet number $y$ for each scalar irreducible representation
$\Phi$,
\begin{equation}
\mathscr{L}_{\textrm{Yukawa}}=y\boldsymbol{171}\cdot\boldsymbol{171}\cdot\Phi\,.\label{eq:171-171-phi}
\end{equation}
There are two possible quantum numbers for $\Phi$: it could either
transform as $\overline{\boldsymbol{3876}}$ or $\overline{\boldsymbol{10830}}$
and reference \cite{Ekstedt:2020gaj} considered the first possibility.
But how might the entries of $Y_{U}$, $Y_{D}$ and $Y_{E}$ be generated
from just a single number? The answer is the following. The scalar
$\Phi$ contains several components which transform as $S\equiv\left(\boldsymbol{1},\boldsymbol{1},0\right)$,
$H\equiv\left(\boldsymbol{1},\boldsymbol{2},1/2\right)$ and $\widetilde{H}\equiv\left(\boldsymbol{1},\boldsymbol{2},-1/2\right)$
under the Standard Model gauge group, and it is the ratio of
their VEVs that control the entries of $Y_{U}$, $Y_{D}$ and $Y_{E}$.
In fact, the coupling constant $y$ in equation (\ref{eq:171-171-phi}) 
merely acts as an overall normalization factor for these three matrices.

In order to go further, we need to break down the single $SU(19)$-invariant
term above into several pieces which are only invariant under the
$G_{SM}$ subgroup. Fortunately, only a few of them are relevant at low
energies. Among others, we find the following $G_{SM}$ representations
inside the larger $SU(19)$ ones:
\begin{align}
\boldsymbol{171} & \ni Q_{i},Q^{c},u_{i}^{c},u,d_{i}^{c},d_{5}^{c},d_{1},d_{2},L_{i}^{c},L_{5}^{c},L_{1},L_{2},e_{i}^{c},e,N_{ij}^{c}\,,\nonumber \\
\overline{\boldsymbol{3876}} & \ni S_{L},S_{DL},S_{D},S_{UD}^{i},S_{QDL}^{i},S_{EL}^{i},S_{N},H_{QN}^{ij},H_{N,i},\widetilde{H}_{DE}^{i},\widetilde{H}_{D}^{ij},\widetilde{H}_{E}^{ij}\,.
\end{align}
The quantum numbers associated to the symbols appearing here have
been mentioned before except those of $Q^{c}$, $u$, $d$, $L^{c}$
and $e$ which are the vector-like partners of the Standard Model
fermion representations. On the other hand the indices $i$ and $j$
indicate that a field transforms under an extra $SU(4)_{F}$ found
inside $SU(19)$, and which commutes with $G_{SM}$. For example, $Q_{i}$
is a quadruplet of this group (lower index), the scalar $S_{EL}^{i}$
is an anti-quadruplet (upper index) and $Q^{c}$ is a singlet (no
index). These indices must be antisymmetrized, therefore $N_{ij}^{c}$
transforms as the antisymmetric product of two quadruplets of $SU(4)_{F}$
--- a sextet.

We then see that there are 4 $Q$'s and 1 $Q^{c}$; 5 $d^{c}$'s and
2 $d$'s, and so on. All in all, there is always an excess of 3 copies
of a fermion representation $X$ over its vector-like partner $X^{c}$
which ensures that at low energies only the Standard Model fermions
are observed. The precise composition of these light fermions is controlled
by the VEVs of the $S$ scalars: indeed we find that the $SU(19)$-invariant
term in equation (\ref{eq:171-171-phi}) contains the $G_{SM}$-invariant
pieces\arraycolsep=1.0pt
\begin{align}
y^{-1}\mathscr{L}_{\textrm{Yukawa}} & \supset Q_{i}\boldsymbol{\mathsf{M}}_{\boldsymbol{\mathsf{Q}}}Q^{c}+u_{i}^{c}\boldsymbol{\mathsf{M}}_{\boldsymbol{\mathsf{U}}}u+e_{i}^{c}\boldsymbol{\mathsf{M}}_{\boldsymbol{\mathsf{E}}}e+N_{ij}^{c}\boldsymbol{\mathsf{M}}_{\boldsymbol{\mathsf{N}}}N_{kl}^{c}+\left(\begin{array}{c}
d_{i}^{c}\\
d_{5}^{c}
\end{array}\right)^{T}\boldsymbol{\mathsf{M}}_{\boldsymbol{\mathsf{D}}}\left(\begin{array}{c}
d_{1}\\
d_{2}
\end{array}\right)+\left(\begin{array}{c}
L_{i}\\
L_{5}
\end{array}\right)^{T}\boldsymbol{\mathsf{M}}_{\boldsymbol{\mathsf{L}}}\left(\begin{array}{c}
L_{1}^{c}\\
L_{2}^{c}
\end{array}\right)
\end{align}
where
\begin{gather}
\boldsymbol{\mathsf{M}}_{\boldsymbol{\mathsf{Q}}}=\frac{1}{3}S_{QDL}^{i}\,,\;\boldsymbol{\mathsf{M}}_{\boldsymbol{\mathsf{U}}}=\frac{\sqrt{2}}{3}S_{UD}^{i}\,,\;\boldsymbol{\mathsf{M}}_{\boldsymbol{\mathsf{E}}}=\sqrt{\frac{2}{3}}S_{EL}^{i}\,,\;\boldsymbol{\mathsf{M}}_{\boldsymbol{\mathsf{N}}}=\frac{1}{4}\sqrt{\frac{2}{3}}\epsilon_{ijkl}S_{N}\,,\\
\boldsymbol{\mathsf{M}}_{\boldsymbol{\mathsf{D}}}=\frac{\sqrt{2}}{3}\left(\begin{array}{cc}
-S_{QDL}^{i} & \sqrt{2}S_{UD}^{i}\\
-S_{DL} & \sqrt{2}S_{D}
\end{array}\right)\,,\;\boldsymbol{\mathsf{M}}_{\boldsymbol{\mathsf{L}}}=\frac{1}{\sqrt{3}}\left(\begin{array}{cc}
-S_{QDL}^{i} & \sqrt{2}S_{EL}^{i}\\
-\frac{2}{\sqrt{3}}S_{L} & S_{DL}
\end{array}\right)\,.
\end{gather}
For example, if $\left\langle S_{QDL}\right\rangle =\left(0,1,0,0\right)^{T}$
and $\left\langle S_{UD}\right\rangle =\left(1,0,0,0\right)^{T}$
then the fermions $Q_{1,3,4}$ are massless until the electroweak
symmetry is broken, and so are $u_{2,3,4}^{c}$. Furthermore, the
right-hand side of equation (\ref{eq:171-171-phi}) also contains the following
interactions with $H$ and $\widetilde{H}$ fields:
\begin{align}
y^{-1}\mathscr{L}_{\textrm{Yukawa}} & \supset Q_{i}\boldsymbol{\mathsf{Y}}_{\boldsymbol{\mathsf{U}}}u_{j}^{c}+Q_{i}\boldsymbol{\mathsf{Y}}_{\boldsymbol{\mathsf{D}}}\left(\begin{array}{c}
d_{j}^{c}\\
d_{5}^{c}
\end{array}\right)+\left(\begin{array}{c}
L_{i}\\
L_{5}
\end{array}\right)^{T}\boldsymbol{\mathsf{Y}}_{\boldsymbol{\mathsf{E}}}e_{i}^{c}+\left(\begin{array}{c}
L_{i}\\
L_{5}
\end{array}\right)^{T}\boldsymbol{\mathsf{Y}}_{\boldsymbol{\mathsf{N}}}N_{jk}^{c}
\end{align}
where $\boldsymbol{\mathsf{Y}}_{\boldsymbol{\mathsf{U}}}=-2/3H_{QN}^{ij}$ and
formulas for the remaining $\boldsymbol{\mathsf{Y}}_{\boldsymbol{\mathsf{X}}}$
can be found in \cite{Ekstedt:2020gaj}. Recall that the two upper
indices mean that $H_{QN}^{ij}$ is a sextet of the $SU(4)_{F}$ flavor
group and therefore it can be written as an anti-symmetric 4 by 4
matrix. The Standard Model Higgs boson $H_{SM}$ must be a combination
of all the $H$ and $\widetilde{H}$ fields, so we can write that
$H_{QN}^{ij}=\Lambda_{QN}^{ij}H+\cdots$ where $\Lambda_{QN}$ is
some anti-symmetric matrix of coefficients. With this notation and
taking $\left\langle S_{QDL}\right\rangle =\left(0,1,0,0\right)^{T}$
and $\left\langle S_{UD}^{i}\right\rangle =\left(1,0,0,0\right)^{T}$ as an example,
the SM $Y_{U}$ matrix would be given by the expression
\begin{equation}
Y_{U}=-\frac{2}{3}y\left(\begin{array}{ccc}
\Lambda_{QN}^{12} & \Lambda_{QN}^{13} & \Lambda_{QN}^{14}\\
-\Lambda_{QN}^{23} & 0 & \Lambda_{QN}^{34}\\
-\Lambda_{QN}^{24} & -\Lambda_{QN}^{34} & 0
\end{array}\right)\,.
\end{equation}

Similar calculations can be done for the remaining Standard Model
Yukawa matrices, as well as for the neutrino masses $1/2\left(m_{\nu}\right)_{\alpha\beta}\nu_{L,\alpha}\nu_{L,\beta}$.
In this way, flavor might be generated effectively at low energies
from a fundamental theory which is flavorless. Still, in order to
confront this model with the observed fermion masses and mixing data,
it would be necessary to also examine the VEVs which minimize the
scalar potential.

\section{Conclusions}

Grand unified theories, which have been proposed and studied for more
than four decades, provide a potential explanation for the Standard
Model quantum numbers, as well as the values of the three gauge couplings.
However, standard GUTs do not explain the phenomena of fermion family
replication. In this work I discussed how non-standard GUTs might
do so.

\section*{Acknowledgments}

I am grateful to Andreas Ekstedt and Michal Malinsk\'y for their collaboration
in the paper \cite{Ekstedt:2020gaj}, on which the present work is
partially based. I acknowledge the financial support from the Grant
Agency of the Czech Republic (GA\v{C}R) through contract number 20-17490S
and from the Charles University Research Center UNCE/SCI/013.

%\bibliographystyle{bibliography/t1}
%\bibliography{bibliography/references}

\end{document}